\documentclass[prb,preprint]{revtex4}
\pdfoutput=1
\usepackage{graphicx}
\begin{document}


\title{Excess oxygen in $\mathbf{La_{2-x}Sr_xCoO_4}$
\medskip }

\date{December 22, 2017} \bigskip

\author{Manfred Bucher \\}
\affiliation{\text{\textnormal{Physics Department, California State University,}} \textnormal{Fresno,}
\textnormal{Fresno, California 93740-8031} \\}

\begin{abstract}
The amount of non-stoichiometric oxygen in $La_{2-x}Sr_xCoO_{4+y}$ is calculated from $Sr$ doping, $x$, and the measured stripe incommensurability $2\delta$ for $x = 1/4$ and $x = 1/3$ based on a linear or square-root dependence of $\delta(x, y)$. The results favor the square-root dependence and indicate that the range of excess oxygen extends to $x < 0.4$ instead of to $x < 0.3$ as previously thought, coinciding with the low-temperature orthorhombic (LTO) phase.

\end{abstract}

\maketitle

Pristine lanthanum cobaltate, $La_2CoO_4$, is an antiferromagnetic Mott insulator. When doped with divalent $Sr$ ionized lanthanum atoms, $La \rightarrow La^{3+} + 3e^-$, are substituted by ionized $Sr \rightarrow Sr^{2+} + 2e^-$. This causes electron deficiency (hole doping) of concentration $n_h = x$. 
If the $Sr$-doped compound contains excess oxygen, denoted as $La_{2-x}Sr_xCoO_{4+y}$,
each such oxygen atom ionizes, $O \rightarrow O^{2-} - \, 2 e^- =  O^{2-} + \, 2 e^+$ , by capturing two electrons from neighboring crystal atoms, leaving two holes behind. The combined concentration of holes from $Sr$ doping and excess oxygen is then
\begin{equation}
p \equiv n_h = x + 2y \, .
\end{equation}

The doped holes destroy magnetic dipole moments in the host crystal, thereby weakening the magnetic interaction. Three-dimensional antiferromagnetism (3D-AFM) collapses when hole doping is at the N\'eel concentration, $p^N(T)$, with a value of $p_0^N \approx 0.25$ at temperature $T=0$. 
With more hole doping glassy 2D-AFM persists and magnetic density waves (MDWs) appear, together with charge-density waves (CDWs). The density waves are incommensurable with the crystal lattice and therefore also called ``stripes.'' Interest in stripes arises from their possible connection with superconductivity in the cuprates and with the enigmatic pseudogap phase.\cite{1} 

An unresolved issue is the doping dependence of the stripes' incommensurability, $\delta(p)$.  Motivated by early experiments  with neutron scattering on lanthanum cuprates,\cite{2} $La_{2-x}Sr_xCuO_4$, the notion arose widely that the stripes' incommensurability depends \emph{linearly} on hole doping,
\begin{equation}
2 \delta(p) = p \,, 
\end{equation}
sometimes called the ``Yamada rule.'' Later experiments with neutron or X-ray scattering in extended doping ranges of $La_{2-x}Ae_xCuO_4$ ($Ae = Sr, Ba$) and $La_{2-x}Sr_xNiO_4$
showed deviations of the incommensurability of MDWs and CDWs from linearity.\cite{3} 
An alternative version, derived in partitioning the $CuO_2$ plane by the doped holes,\cite{4} has a \emph{square-root} dependence,
\begin{equation}
2 \delta(p) = \sqrt{p - p_0}  \, ,
\end{equation}
with an off-set by the N\'eel concentration, $p_0 = 0.25 \approx p_0^N$, in the radicand. Neutron scattering experiments\cite{5} on $La_{2-x}Sr_xCoO_4$ in the doping range $0.4 \leq x \leq 0.5$ (see Fig. 1) gave values of stripe incommensurability that are equally well described by a linear or square-root doping dependence, Eqs. (2) or (3), leaving the controversy unresolved.

\pagebreak
Recent experiments with neutron scattering at lower doping,\cite{6,7,8} nominally at $x = 1/4$ and $x = 1/3$, give stripe incommensurabilities close to linear doping dependence, Eq. (2), listed in Table I. Here ``nominal'' means by weight. In the case of nominally $x = 1/4$ the $Sr$ content was also determined by both X-ray analysis and electron probe microanalysis   

\bigskip

\includegraphics[width=6in]{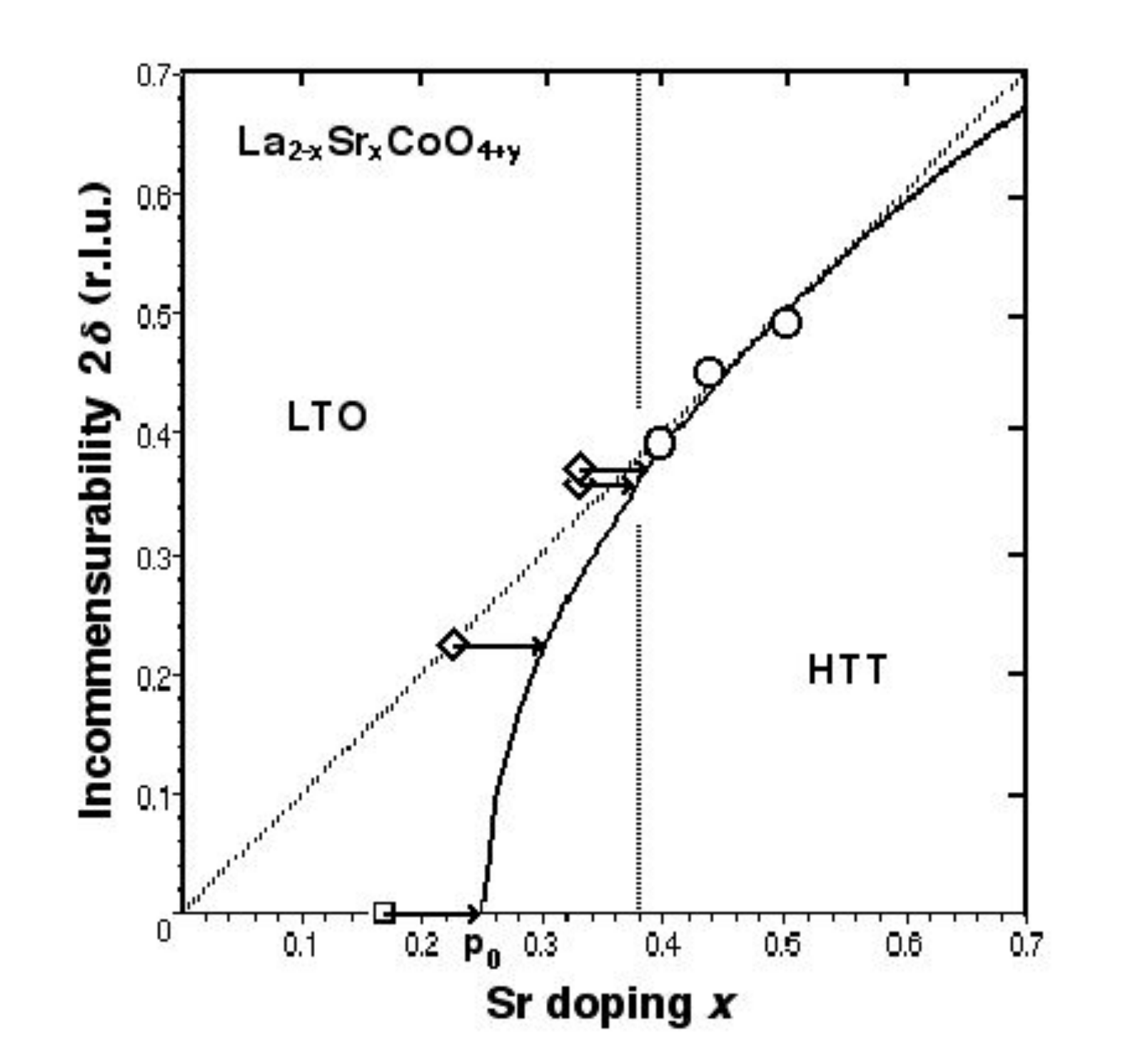}

\footnotesize 
\noindent FIG. 1. Doubled incommensurability, $2 \delta$, of magnetic stripes in $La_{2-x}Sr_{x}CoO_{4+y}$ in dependence on $Sr$ doping $x$. Circles (Ref. 5) and diamonds (Refs. 6 - 8) show experimental data from neutron scattering. 
The $Sr$ doping at the onset of stripes, shown by the square symbol, is inferred from muon-spin rotation experiments (Ref. 9). 
The vertical dotted line (Ref. 5) separates the high-temperature tetragonal (HTT) phase (here extrapolated to $T = 0$) where stoichiometry exists and the low-temperature orthorhombic (LTO) phase where excess oyygen is present. The solid curve represents a square-root dependence on hole density, $2\delta(p)$, Eq. (3). The hatched diagonal represents a linear dependence of  $2\delta(p)$, Eq. (2). The arrows show the contribution to hole doping from excess oxygen of concentration $y$.
\normalsize

\noindent yielding values of $x = 0.22$ and $x = 0.238$, respectively, listed by their average $x = 0.23 \pm 0.01$ in Table I. 
Likewise the $Sr$ content of some $x = 1/3$ samples was determined by X-ray analysis.\cite{8} The most relevant value is listed in Table I.
Both Ref. 6 and 7 note indications of excess oxygen, believed to approach stoichiometry ($y=0$) at $x = 0.3$.\cite{6,7} 
Excess oxygen of $y \approx 0.01$ has been inferred from the deviation of $2\delta$ from $x$, Eq. (4).\cite{8} Independently, excess oxygen at low $x$ has also been inferred from the suppression of the N\'{e}el temperature $T_N(x)$.\cite{9} 

From the value of $Sr$ doping, $x$, and the measured stripe incommensurability, $2\delta$, the concentration of excess oxygen, $y$, can be determined according to the doping dependence $\delta(p)$. For a linear doping dependence a combination of Eqs. (1) and (2) gives
\begin{equation}
2y_{lin} = 2 \delta - x\, . 
\end{equation}
For a square-root doping dependence a combination of Eqs. (1) and (3) gives
\begin{equation}
2y_{sqrt} = (2\delta)^2 + 0.25-x \, . 
\end{equation}

Table I shows that the calculation with linear doping dependence, Eq. (3), gives excess oxygen of $\sim1$ \% or $\sim2$ \% for $x = 1/3$ but 0.6 \% oxygen \emph{deficiency} for $x = 1/4$. Such oxygen deficiency is in disagreement with the indications for excess oxygen\cite{6,7,8,9} for low $Sr$ doping, and also with the trend of decreasing oxygen excess $y$ with increasing $x$. In contrast, the calculation based on a square-root dependence of $\delta(p)$, Eq. (4), gives 3.5 \% oxygen excess for $x = 1/4$, decreasing to $ \sim 2.4$ \% at increased doping $x = 1/3$. 
A comparison of Eqs. (4) and (5) gives $y_{sqrt} > y_{lin}$ for any $Sr$ doping $x$ and incommensurability $2\delta \ne 0.5$. Accordingly, for given values of $x$ and $2\delta$  excess of oxygen is larger for the square-root doping dependence than for the linear one. This means that if the square-root dependence of $\delta(p)$ is valid, then 

\medskip

\begin{table}[h!]
\begin{tabular}{| p{2.1cm}|p{2.1cm}|p{2.2cm}|p{1.1cm}|p{1.1cm}|p{0.7cm}|  }
 \hline  \hline
  $x$ nominal &  $x$ measured &  $2 \delta$ measured & $y_{lin}$ & $y_{sqrt} $ & Ref. \\
 \hline 
$1/4 = 0.25$ &  $0.23 \pm 0.01$  &  $0.224 $ & -0.006 & 0.035 & 6\\
$1/3 = 0.333$ &  &  $0.37 \pm 0.01 $ & 0.018 & 0.027  & 7\\
$1/3 = 0.333$ & 0.332 &  0.353  & 0.01 & 0.021  & 8\\
 \hline   \hline
\end{tabular}
\caption{Nominal and measured $Sr$ content $x$ in $La_{2-x}Sr_xCoO_{4+y}$, measured stripe incommensurability $2\delta$, and excess oxygen content $y_{lin}$ and $y_{sqrt}$ calculated with linear or square-root doping dependence, Eqs. (4) or (5), respectively.}

\label{table:1}
\end{table}

\pagebreak
\noindent oxygen excess is larger than expected for linear $\delta(p)$.

The earlier finding\cite{3} that the measured incommensurability\cite{5} $2\delta$ for $0.4 \leq x \leq 0.5$ closely agrees with both Eq. (2) and (3) [using $y = 0$ in Eq. (1)], indicates that stoichiometry has been reached in that doping range. 
It is corroborated by confirmed stoichiometry in $La_{1.5}Sr_{0.5}CoO_4$.\cite{10}
However, if the square-root dependence of $\delta(p)$ is valid, Eq. (3), then the present findings suggest that the range of oxygen non-stoichiometry extends to $x < 0.4$ instead of to $x < 0.3$ as previously thought.\cite{6,7,9}
In the $La_{2-x}Sr_xCoO_4$ phase diagram the  boundary of the high-temperature tetragonal (HTT) and low-temperature orthorhombic (LTO) phases, extrapolated to $T = 0$, meets the doping axis at $x \approx 0.4$.\cite{11}
The inference of excess oxygen in the LTO phase but stoichiometry in the HTT phase is consistent with the present findings and provides a nice explanation\cite{12} for the sudden change from non-stoichiometry to stoichiometry. 

Muon-spin rotation experiments indicate that incommensurate stripes are present in the (nominal) doping region $\sim 0.2 \leq x \leq 0.6$.\cite{9} 
A close inspection of the data\cite{13} shows that the asymmetry $\overline{A}_0(x)$ of muon-spin polarization, the relaxation rate $\lambda(x)$, and the N\'{e}el temperature $T_N(x)$ derived from $\overline{A}_0$, all steeply rise with a leading edge at $x_0 = 0.17$. This can be associated with the onset of MDWs where $2\delta = 0$ holds. Using  these values Eq. (5) gives excess oxygen of $y_{sqrt} = 0.04$ at the onset of stripes---a value consistent with the trend of $y_{sqrt}$ in Table I. The corresponding contribution to  hole density is shown in Fig. 1 by the arrow from the square symbol to the foot of the square-root curve.

\bigskip \bigskip

\centerline{ \textbf{ACKNOWLEDGMENT}}
\noindent I thank Paul Freeman for recent literature references and valuable correspondence.

\end{document}